\documentclass[aip,jcp,preprint,twocolumn]{revtex4-1}

\usepackage{amsmath}
\usepackage{amsfonts}
\usepackage{amssymb}
\usepackage{graphicx}

\begin{document}

\title{Investigation of the effect of surface phosphate ester dispersant on viscosity by coarse-grain modeling of BaTiO$_3$ slurry}

\author{Hiroya Nakata}
\email{hiroya.nakata.gt@kyocera.jp}
\affiliation{Kyocera Corporation, Research Institute for Advanced Materials and Devices, 3-5-3 Hikaridai Seika-cho Soraku-gun Kyoto 619-0237, Japan.}
\author{Takayoshi Kiguchi}
\affiliation{Department of Chemical Science and Technology, Faculty of Bioscience and Applied Chemistry, Hosei University, 3-7-2, Kajino-cho, Koganei,Tokyo,184-8584, Japan.}
\author{Osamu Hino}
\email{ohino@x-ability.jp}
\affiliation{X-Ability Co., Ltd.,Ishikawa Building, 3rd floor, Hongo 4-1-5 Bunkyo-ku, Tokyo 113-0033, Tokyo, +81-3-5800-7731}

\begin{abstract}
  To understand the role of phosphate ester dispersant, we investigated the rheology of a BaTiO$_3$ slurry.
  For the model case, a coarse-grain molecular 
  dynamics (CGMD) simulation was performed 
  with the butyral polymer didodecyl hydrogen phosphate (DHP), in the toluene/ethanol solvent. 
  By systematically analyzing the effect of DHP from an atomic-scale first principle and from all-atom MD to micro-scale CGMD
  simulation, we investigated how the adsorption of a DHP dispersant on a BaTiO$_3$ surface affects the microstructure rheology of a BaTiO$_3$ slurry.
  The first-principle and all-atom MD simulation suggests
  that DHP molecules prefer to locate near the BaTiO$_3$ surface. 
  CGMD simulation shows a reduction in viscosity with an increase in dispersants, suggesting 
  that the dispersant population near the BaTiO$_3$ surface plays a key role 
  in controlling the rheology of the BaTiO$_3$ slurry.
  In this study, we propose an approach for understanding the BaTiO$_3$
  slurry with molecular-level simulations, which would be
  a useful tool for efficient optimization of slurry preparation.
\end{abstract}


\maketitle



\section{Introduction}
  With the development of electronic equipment, including mobile phones and that for electric vehicles, and the rapid increase in its 
  production quantities in the past several decades, multilayer ceramic capacitors (MLCCs) play an increasingly important role in today's electronics industry
\cite{kishi2003base,hong2019perspectives}. 
  At the same time, the higher performance of such devices and equipment has increased the demand for miniaturization and high volumetric 
  capacitance density and reliability of MLCCs. 
  MLCCs are made by alternately stacking numerous ceramic dielectric layers and metal electrodes, 
 and significant effort is invested not only in investigating the effects of the component\cite{sakabe2002dielectric,gong2012electrical,li2021colossal} 
 and grain size\cite{buscaglia2020size,zhu2021grain} of the BaTiO$_3$-based 
   material on the dielectric properties, but also in improving the process of producing the thin layers. 
  As a result, the dielectric layer has now reached a thickness of less than a few micrometers 
  in the latest MLCCs\cite{buscaglia2020size,suzuki2020effect,okuma2021microstructural,suzuki2021suppressive}. 
   In these very thin MLCCs, highly controlled microstructure homogeneity is required for high reliability and 
   improved manufacturing yield. For instance, microstructure heterogeneity has been reported to cause degradation of MLCCs; 
   by visualizing electrically degraded areas in MLCCs, Sada et al.\cite{sada2017analysis} showed that a thinner dielectric layer caused early failure. 
   Nagayoshi et al.\cite{nagayoshi2020analyses} reported that some large grains were observed in the degraded areas. 
   Samantaray et al.\cite{samantaray2012electrode,samantaray2012electrode02} showed by the finite element method th
   at electrode porosity and roughness in MLCCs could lead to a higher local electric field and 
   leakage current. 
   Heath et al.\cite{heath2019electric} calculated the electric field at the interface between the dielectric layer 
    and the electrode and showed that the largest enhancement of field occurred in the sharper-pointed part of the electrode. 
    In the production of MLCCs, the ceramic particles used as the raw materials of the dielectric layer are usually dispersed in organic solvent. 
    Thus, it is important that the slurry be made to be well dispersed, stable and uniform and have appropriate rheological characteristics for tape casting, 
    so that the green sheets are sufficiently homogeneous and smooth\cite{chu2008dispersion}. 
    Accordingly, in general, a dispersant, binder, plasticizer and deformer are added to the slurry. 
    However, the appropriate amount of each organic additive depends on the kind of such organic additives and media, and their combination, 
     making slurry preparation optimization necessary\cite{mikeska1988non,bhattacharjee1993polyvinyl,tseng2003effect,iwata2019effect}. 

    With recent improvements in simulation performance,  theoretical analysis by numerical simulation has become one of the trends for 
    reducing this optimization difficulty. 
    Enabling the creation of a computational connection between the rheological characteristic of the slurry and the dispersion and 
    aggregation states of the particles would be useful, because slurry is often practically evaluated by measuring viscosity, 
    which is fast and 
    easy to perform\cite{chu2008dispersion,mikeska1988non,bhattacharjee1993polyvinyl,tseng2003effect,iwata2019effect,dong2017optimization,iwata2020determination,ding2020optimization}.
    The adsorption of dispersant for various particles was also studied by  molecular dynamics (MD) simulations and density functional theory (DFT) calculations because
the behavior of dispersant close to the solid surface plays a major role in predicting the rheological characteristic of a slurry\cite{chen2007adsorption,zhao2017effect,chun2021first}. 
    Numerous  experiments  and theoretical simulations have  been  performed for the optimization of slurry preparation,
    but a  molecular-level  exploration  of the microstructure--rheology  relationship is  still  lacking. 
    The main obstacle in connecting molecular-level surface interaction and the microstructure--rheology relationship
    is the limitation of computational time of the atomistic simulation.   
    
    To bridge the molecular-level surface interaction to microstructure rheology,
    a CG model\cite{grest1986molecular,kremer1990dynamics} 
    provides a significantly large and long time-scale simulation with 
    reasonable computational cost. 
    There are two main approaches for constructing the CG model.
    The top-down approach fits the CG model to reproduce experimental data, whereas the 
    bottom-up approach fits the interaction force and parameter to reproduce the configuration 
    of an atomistic MD simulation\cite{trement2014conservative,reith2003deriving}.  
    The CG model is applied to understand the friction\cite{kreer2001frictional}, 
    polymer melt\cite{padding2002time,spyriouni2007coarse,guerrault2004dissipative,harmandaris2005molecular}, 
    viscosity\cite{wang2019implicit,maurel2012multiscale,li2012nanoparticle,wendt2019effect} and 
    many other properties\cite{ghoufi2012coarse,hagita2019two,harton2010immobilized,qian2008temperature}.    
    However, application of the CG model to the solid--polymer interface is still challenging,
    because there is neither a universal parameter for the atomistic simulation of the polymer--solid interface, nor 
    detailed experimental information. 
    Thus, both top-down and bottom-up approaches have difficulty generating a CG model of the polymer--solid
    interface.  Due to the difficulty in applying the CG model to the polymer--solid interface,
   the CG model has been applied to only a few types of systems such as
    silica\cite{ndoro2011interface,ghanbari2012interphase,bogoslovov2008effect,ghanbari2012interphase,ndoro2012interface}, 
    alumina\cite{kacar2013structure}, 
    gold\cite{johnston2013hierarchical}, 
    and graphene or graphite\cite{eslami2013thick,daoulas2005detailed,pandey2014multiscale}.
    To the best of our knowledge, no CG study has applied the method to aBaTiO$_3$-based 
    slurry to understand the relationship between molecular-level surface interaction and 
    microstructure rheology. 
    
    The purpose of this study was to fill the gap between molecular-level surface interaction 
    and the microstructure rheology of a BaTiO$_3$-based slurry by a systematic 
    analysis from atomistic scale  molecular adsorption to micro-scale rheology. 
    The present work presents a systematic approach that connects the different level and scale of simulations
    by using quantum-level DFT, all-atom MD, and CGMD simulations. 
    First, the adsorption energy of a dispersant on a BaTiO$_3$ surface is simulated with a DFT simulation, generating 
    a force-field parameter for all-atom MD simulation. The distributions of the dispersant, binder and  
    solvent near the BaTiO$_3$ surface are then evaluated to understand how the dispersant affects the structure of the
    interface. Finally, by using the distributions on the BaTiO$_3$ interface, a CGMD force-field parameter 
    is prepared, and the model is applied to predict the rheology of the BaTiO$_3$-based slurry
    to understand how the local adsorption of dispersant affects the rheology.

\section{Theoretical method and computational detail}

\begin{figure}
          \begin{center}
           \includegraphics[clip,width=8.0cm]{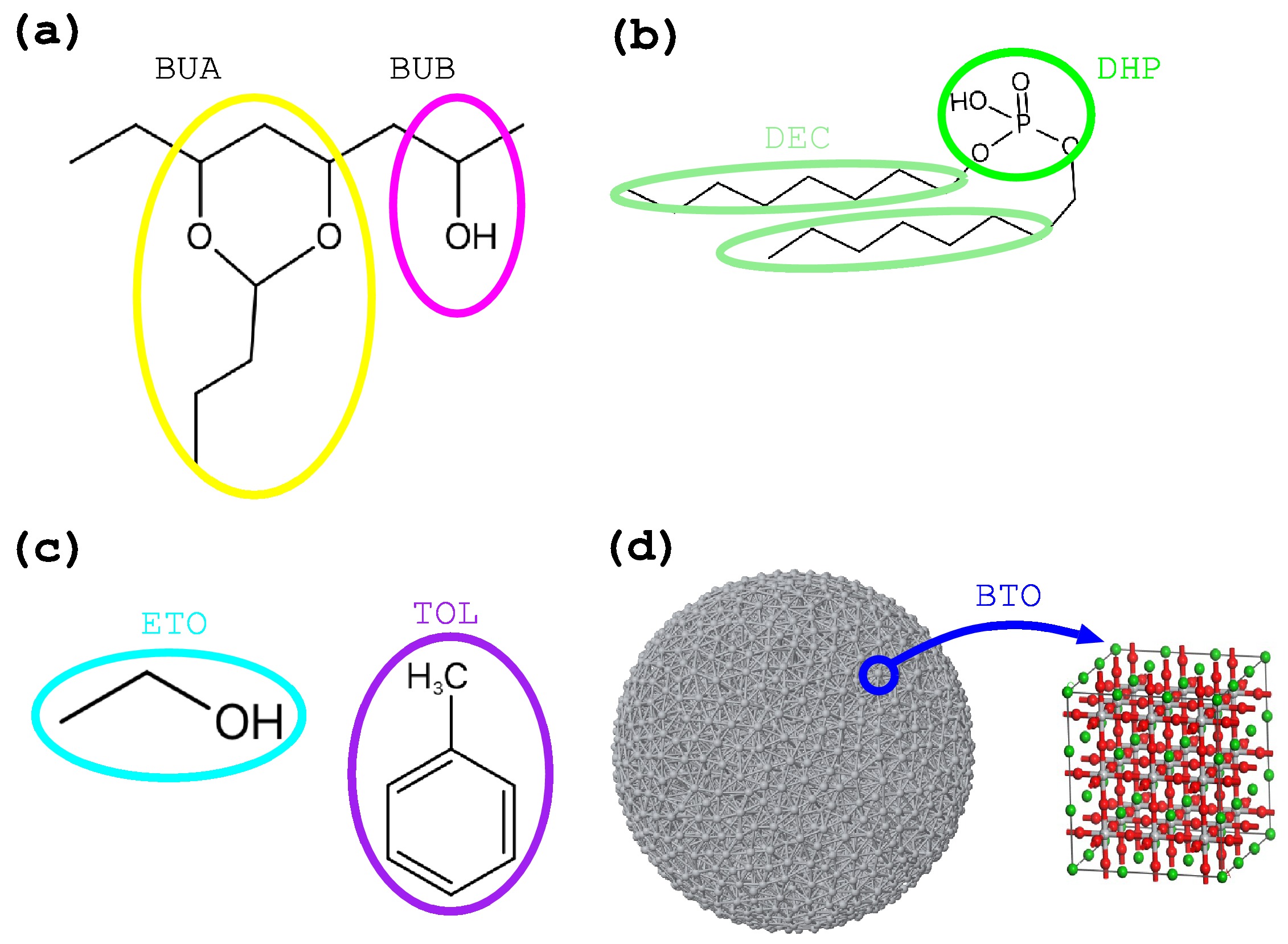} \\
          \end{center}
	      \caption{
              Chemical formulation and crystal structure used in this study:
              (a) Butyral polymer,
              (b) didodecyl hydrogen phosphate (DHP),
              (c) solvent molecules (ethanol and toluene),
              (d) BaTiO$_3$ particles.
              \label{Figure01}
		      }
	\end{figure}

    To investigate the relationship between the molecular-level surface interaction 
    and the microstructure rheology of a BaTiO$_3$-based slurry, one of the most popular 
    combinations of binder,  dispersant and solvent was selected as  the model case\cite{vinothini2006optimization}.  
    The molecular structures and compositions used in our simulation are listed 
    in Figure~\ref{Figure01} and Table~\ref{TABLE01}.
    The butyral polymer is used for binder, and the molecular structure is shown in 
    Figure~\ref{Figure01}-(a) 
    (Throughout this study, we denote the molecular structure as BUA and BUB for simplicity).
    As for the dispersant,  didodecyl hydrogen phosphate (DHP) was used (Figure~\ref{Figure01}-(b)).
    As for the solvent molecules,
    we used toluene (TON) and ethanol (ETO) as solvent of the BaTiO$_3$ slurry (Figure~\ref{Figure01}-(c) and (d)).
    For BaTiO$_3$ particles, a cubic crystal structure was used (Figure~\ref{Figure01}-(d)).
    In the CGMD simulation model, the group of atoms in a molecule are simplified into one particle.
    The correspondence between the molecular structure and simplified particle used in the CGMD simulation
    is listed in the colored open circles in Figure~\ref{Figure01}. 
    To construct the CG model for BaTiO$_3$, we treat the BaTiO$_3$ particle as a rigid sphere (Figure~\ref{Figure01}-(d)), 
    with the radius set at 2, 8 and  15 nm, respectively. The rigid sphere is then divided into 
    a mesh, and the contact points are treated as the interaction sites for CG model.

    To systematically analyze the role of the DHP dispersant on the rheology of the slurry, 
    three types of analysis were performed.
    First, the interaction energies between the BaTiO$_3$ surface and the respective molecules were 
    estimated with the first-principle simulation.
    Because the first-principle simulation is limited relatively small size, it can take account only 
    the change of potential energy with single z-direction (1D).
    In order to consider the 3D effect (such as molecular rotation and translation), 
    it is necessary to perform MD simulation.
    Thus, secondly, 
    we performed an all-atom MD simulation  to statistically analyze the structure of the interface of the BaTiO$_3$ slurry
    by generating a force field to reproduce the first principle interaction energies.
    For this purpose, we evaluated the distribution of each type of molecule near the BaTiO$_3$ [001] surface 
    terminated with OH.
    The size of simulation model in all-atom MD is still limited, and all-atom MD can take account only the molecular distribution
    close to the solid surface. Thus it is necessary to extend the scale of simulation to evaluate rheology of 
    slurry.
    Finally, from the obtained distributions of the respective molecules near the BaTiO$_3$ surface, 
    the Boltzmann inversion method\cite{reith2003deriving,guerrault2004dissipative} was used to obtain 
    the parameter of the CG model (See detail in section II-D), and we evaluated
    how the interface geometry affects the rheology of the BaTiO$_3$ slurry.

\subsection{First-principle simulation to investigate the interaction energy of the BaTiO$_3$ surface}
  A first-principle simulation (DFT) was performed using the
  Quantum Espresso software\cite{qe01,qe02} with a
  Perdew--Burke--Ernzerhof functional\cite{PBE01,PBE02}.  
  We used ultrasoft pseudopotentials\cite{ultrasoft}
  and  the cutoff energy for the plane-wave basis set was taken to be 300 Ry. 
  A default convergence criterion of  1.0D-4 a.u  
  was used for the geometry optimizations, 
  and 1.0D-6 a.u  
  for the self-consistent field calculations of the electronic states.

  First, geometry optimizations were performed to determine the interaction energy
  between the BTO surface and the respective molecules (BUA, BUB, DHP, ETO and TON).
  The geometry optimizations were performed with 24 different initial geometries,
  in which we rotated the respective molecules  12 times by 30 degrees around the $x$- and $y$-axes.  
  The most stable geometry was then selected among the 24 different molecular orientations,
  and the potential energy surface (PES) along the $z$ direction was evaluated
  to obtain the interaction energy between the organic  molecules and the BaTiO$_3$ surface.
  For surface model, we used BaTiO$_3$ [001] surface terminated by OH\cite{eglitis2015comparative,wegmann2004xps}.

\subsection{All-atom MD simulation to obtain the distribution of molecules on the BaTiO$_3$ surface}
  All the MD simulations were performed with the LAMMPS software\cite{LammpsPack}
  To evaluate the distribution of respective molecules near the BaTiO$_3$ surface,
  we performed an all-atom MD simulation. 
  For this purpose, three different systems were investigated, in which 
  different compositions of organic bulk models were generated: 
  a) 1000 TON, 1000 ETO and 1000 DHP,
  b) 1500 BUA and  1500 DHP, and 
  c) 1500 BUB and  1500 DHP, respectively.
  To statistically sample the distribution of the interface structure, 
  five different initial geometries were randomly generated  for bulk-organic molecules using packmol\cite{martinez2009packmol},
  and were equilibrated by 1 ns. The final geometry of the MD simulation
   was then used for the organic bulk model. 
  The organic bulk models were merged with the BaTiO$_3$ cubic crystal, generating
  the initial geometry of the interface models. 
   Finally, a 2-ns MD simulation was performed using the interface models,
   and the final 1-ns trajectory was used for molecular distribution analysis.
   The molecular distributions were used to construct the CG model 
   to evaluate the rheology of the BaTiO$_3$-based slurry 
   (See section II-D for detail).

  The force-field parameter for the respective molecules (BUA, BUB, DHP, ETO and TON) was prepared 
  using the general AMBER force field (GAFF), and the atomic charges were estimated 
  with restrained electrostatic potential (RESP). To make the RESP charge,
  DFT with the Becke three-parameter
  hybrid exchange functional combined with the Lee--Yang--Parr correlation
  functional (B3LYP)\cite{cite31,cite32} was used,
  and the 6-31G(d,p) basis set was applied.
  The Quantum Mechanic (QM)  calculation was performed with GAMESS\cite{GAMESS1,GAMESS2,GAMESS3}.
  Note that the nonbonding parameters between the BaTiO$_3$ surface and respective molecules 
  were determined to reproduce the interaction energy changes along the z-axis, 
  which were obtained by the above first-principle simulation 
    (See section III-A for detail).

\subsection{CG  MD simulation for predicting the rheology of the BaTiO$_3$ slurry}
    To understand how the polymer--solid interface affects the rheology of the BaTiO$_3$ slurry,
    three different slurry model compositions were investigated 
    (see TABLE~\ref{TABLE01}).
    The length of polymer chain was set at 100, where the composition of BUA and BUB  was 63 and    
    37, respectively. Thus, the molecular weight of the polymer chain was 10789.
    The CG model for the butyral polymer was generated by a beads-and-spring model\cite{grest1986molecular}.
    To reproduce the geometry of polymer, the length of spring is set 7.0 \AA\ for BUA,
    5.6 \AA\ for BUB, and 6.3 \AA\ between BUA and BUB. Likewise the spring between DHP and DEC is 
    set 5.7 \AA. The force constant of the spring is set 300 kcal/mol.

\begin{table}[h!]
\caption[]{
            Composition of the respective BaTiO$_3$ slurry model; the weight 
            fraction is  shown,
            where BTO, PVB, DHP, TOL and ETO denote BaTiO$_3$, poly-butyral,
            didodecyl hydrogen phosphate, toluene and ethanol.   
\label{TABLE01}}
\begin{tabular}{lrrr}\hline
      &  model01  &   model02 &    model03 \\\hline
 BTO  &   0.30    &    0.20   &     0.10   \\
 PVB  &   0.28    &    0.28   &     0.28   \\
 DHP  &   0.03    &    0.13   &     0.23   \\
 TOL  &   0.24    &    0.24   &     0.24   \\
 ETO  &   0.15    &    0.15   &     0.15   \\\hline
\end{tabular}
\end{table}%

    First, the initial simulation models were randomly generated using packmol\cite{martinez2009packmol}, where 
    the density of the slurry model was set at 0.8 g/cm$^3$, which nearly corresponds to TOL ETO
    solvent densities. A CG MD simulation was then performed until the stable radial distribution function (RDF) was obtained,
    which took about 5 ns of simulation time. The time step for the equilibration is 5 fs.  
     NVT-MD was performed
     with Nose-Hoover's thermostat\cite{nose1984molecular,nose1984unified},
    and the temperature was set to 300 K. 
    The final geometry and velocity were used to obtain the viscosity of the respective models.
    The viscosity was obtained using the SLLOD method implemented in the LAMMPS software\cite{evans1984nonlinear,daivis2006simple}.
     The MD simulation is performed with the SLLOD equation of 
     motion as follows: 
      \begin{align}
       \frac{d \mathbf{r}_i}{dt}
        = &
        \mathbf{v}_i + s r_{i,z}
        \left(
        \begin{array}{cc}
             1  \\
             0  \\
             0
        \end{array}
        \right)
      \\
      \frac{d \mathbf{v}_i}{dt}
        = & 
          \frac{1}{m_i} \mathbf{f}_i - s v_{i,z} 
          \left(
          \begin{array}{cc}
             1  \\
             0  \\
             0
          \end{array}
        \right) 
      \end{align},
      where $i$, $z$, $s$ are particle index, xyz axis,
      and shear rate. 
      In this study, the viscosity was evaluated with several different 
      values of $s$ from 0.1$\times 10^{-5}$ to 5.0$\times 10^{-5}$.
      $\mathbf{r}$, $\mathbf{v}$, and $\mathbf{f}$
      are position, velocity, and force of the respective 
      particles. $\mathbf{m}$ is particle weight.
      Then the viscosity $\eta$ may be calculated from the stress tensor 
      $\mathbf{P}$ as follows:
      \begin{align}
        \eta = - \frac{1}{s} 
                 \langle
                    P_{xz}
                 \rangle
      \end{align}
    To obtain the viscosity of the slurry model, we also performed a 5-ns MD simulation, and a final 1-ns simulation 
    result was used for the viscosity evaluation.

   \subsection{CG pair interaction potential}
    The pair interaction parameters of the CG model were determined using the Boltzmann inversion method\cite{reith2003deriving,guerrault2004dissipative},
    where the potentials between particle type $i$ and $j$
    ($V_{ij,n}$) are determined by RDF ($g_{ij}(r)$ from 
    the all-atom MD simulation:
    \begin{align}
      V_{ij,0} (r)  =&  -k_B T \mathrm{ln}(g_{ij,0}(r))
      \\
       V_{ij,n} (r) =&  V_{ij,n-1} (r)  
                    -   \alpha k_B T 
                        \mathrm{ln}(g_{ij,n-1}(r)).
    \end{align}
      $k_B$, $T$, and $\alpha$ are Boltzmann constant,
      temperature, and scaling factor, and we use 
      $\alpha = 0.5$. $n$ is number of iteration step.
      $g_{ij,0}(r)$ is the RDF estimated by all atom MD 
      simulation, while $g_{ij,n}(r), n=1,2, \cdots $
      is the RDF estimated by CGMD simulation using
      $V_{ij,n}$. Thus, it is necessary to estimate 
      $g_{ij,0}(r)$ for all pair of $i$ and $j$.
      Because we have six different particle types for 
      organic molecules (Figure~\ref{Figure01}), 21 different 
      combination of RDFs are necessary to evaluate with 
      all-atom MD simulation. 
      The all-atom MD models for generating the RDFs
      and the comparison with the RDF of final CGMD simulation 
      is shown in supporting information.

    For the pair interaction between organic molecules, the RDFs for each pair of molecules were estimated by 1 ns MD simulations, and a final 500-ps trajectory was used.
    The pair interaction parameters between BaTiO$_3$ and the molecular interface were determined by the molecular distributions along the z-axis, which were obtained in 
    the  previous subsection (See also section III-B for detail).

\section{Results and discussion}

\subsection{Interaction energy between organic molecules and the BaTiO$_3$ surface}
\begin{figure}
          \begin{center}
           \includegraphics[clip,width=8.0cm]{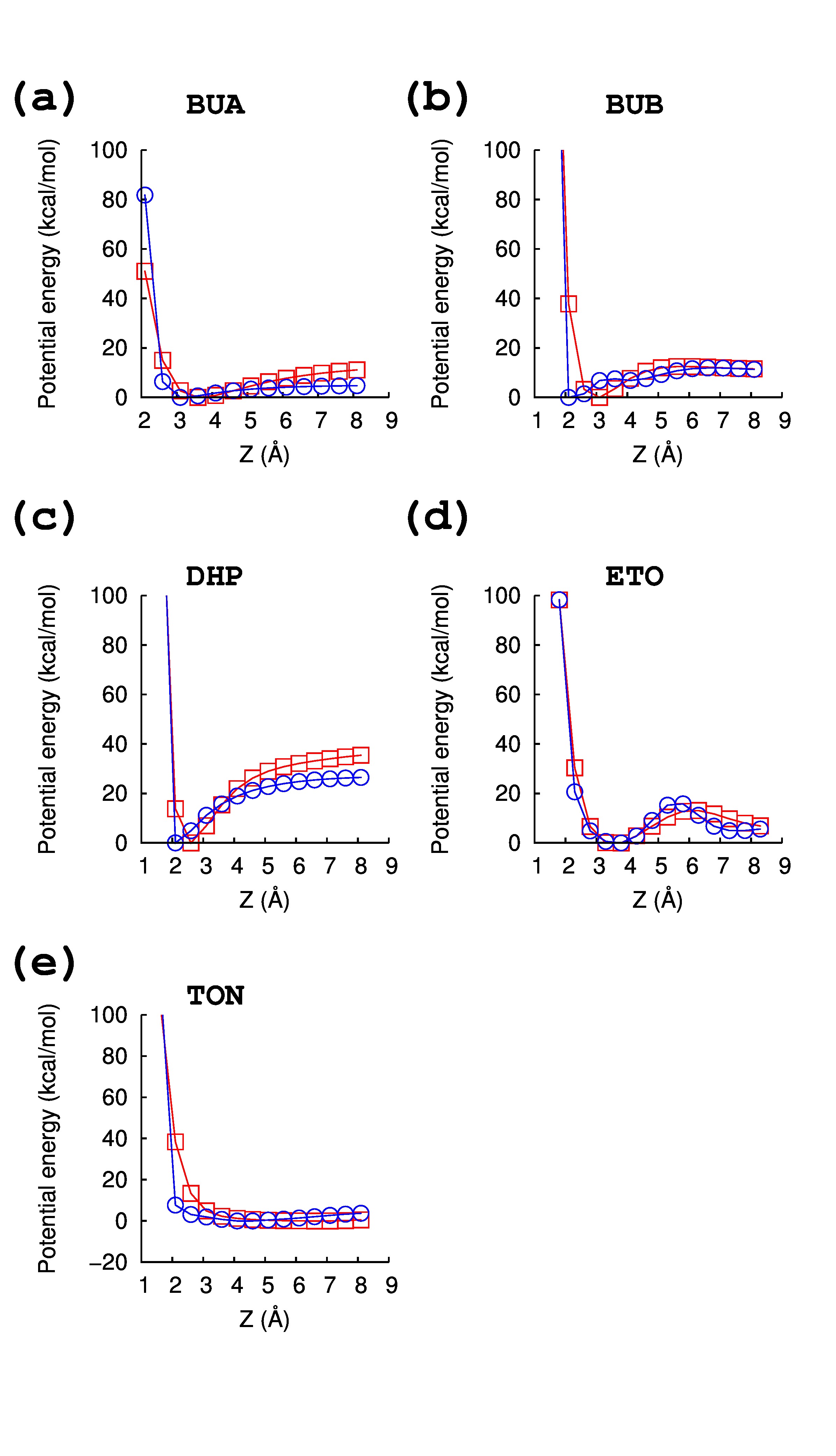} \\
          \end{center}
	      \caption{
               Interaction energy analysis between the BaTiO$_3$ surface and respective molecules.
               Red open squares denote the result of the QM simulation, and 
               blue open circles denote the result of the molecular mechanic simulation.  
              (a) BUA, (b) BUB, (c) DHP, (d) ETO and (e) TON.
              \label{Figure02}
		      }
	\end{figure}

    The adsorption energy of the respective molecules (BUA, BUB, DHP, ETO and TON) 
    was evaluated with the first-principle simulation, and the results are shown in Figure~\ref{Figure02} and TABLE~\ref{TABLE02}. 
     \begin{table}[h!]
     \caption[]{
           Interaction energy between surface BaTiO$_3$ and respective molecules
            (Unit is kcal/mol).
     \label{TABLE02}}
     \begin{tabular}{lrrr}\hline
           &   QM      &      MM*  \\\hline
     BUA   &   8.11    &     4.72  \\
     BUB   &  11.51    &    11.31  \\
     DHP   &  35.44    &    26.46  \\
     ETO   &   6.85    &     5.58  \\
     TON   &   0.53    &     3.74  \\\hline
     * MM denotes Molecular Mechanic.
     \end{tabular}
     \end{table}%
    As shown in TABLE~\ref{TABLE02}, the dispersant molecule (DHP) has the largest interaction energy compared 
    with the other molecules, because the phosphate ester has a strong polarizability.
    In addition, the minimum energy point of DHP was shorter than that of the other molecules, and a longer range attraction force 
    was observed, suggesting that DHP is preferentially located in the vicinity of the BaTiO$_3$ interface. 
    The attraction interaction of DHP is reasonable because the DHP is well known as a strong dispersant  
    for slurry.

    Using the obtained PES, the force-field parameter between BaTiO$_3$ and the respective molecules was 
    determined. The nonbonding interaction energy  between atom $i$ of BaTiO$_3$ and  atom $j$ of the organic molecule
    is evaluated as follows:
    \begin{align}
       E^\mathrm{coul}_{ij} = &  \frac{q_i q_j}{\epsilon r_{ij}}
   \\
       E^\mathrm{vdw}_{ij}  = &  4 \epsilon_{ij}
                                   \left[   
                                     \left(
                                     \frac{\sigma_{ij}}{r_{ij}}
                                     \right)^{12}  
                                   -
                                     \left(
                                     \frac{\sigma_{ij}}{r_{ij}}
                                     \right)^{6}  
                                   \right],  
      \label{nonBonding01}
    \end{align}
    where $q_i$, $r_{ij}$, $\epsilon_{ij}$ and $\sigma_{ij}$ are the 
    atomic charge, distance between atoms and van der Waals parameters. 
    Thus, the necessary parameters to perform an MD simulation are the atomic charge $q_i$, 
    and van der Waals pair interaction $\epsilon_{ij}$ and $\sigma_{ij}$. 
    The parameters $\epsilon_{ij}$ and $\sigma_{ij}$ are estimated by taking the average of 
    the independent parameter of atoms $i$ and $j$ as follows:
    \begin{align}
         \sigma_{ij} =&  
                       \left(
                         \sigma_{i}  + \sigma_{j} 
                       \right) / 2
     \\
         \epsilon_{ij}   =&  
                       \sqrt{
                       \left(
                         \epsilon_{i}  \epsilon_{j} 
                       \right)}.
      \label{nonBonding02}
	\end{align}
    Because  we used the GAFF for the force-field parameter of organic molecules, 
    the parameters $\sigma_{i}$ and $\epsilon_{i}$ for organic molecules  are already reserved
    to describe the interactions between the respective molecules.
    Then, the changeable parameters for surface interaction are $q_i$, $\sigma_{i}$ and $\epsilon_{i}$  
    of BaTiO$_3$ crystal, and these parameters were set to reproduce 
    the PES obtained by the first-principle simulation.  
    The parameters used in the all-atom MD simulation are listed in TABLE~\ref{TABLE03},
    and the PES results estimated by the parameter are indicated in Figure~\ref{Figure02}
    by the blue line and open circles and are in good agreement with the PES evaluated by first principle. 
    We analyzed the molecular distribution 
    near BaTiO$_3$ with the all-atom MD simulation using the obtained force-field parameter.
  
     \begin{table}[h!]
     \caption[]{
           Parameter for the nonbonding interaction energy of surface BaTiO$_3$,
           where Q, $\epsilon$, $\sigma$ denote the charge and  van der Waals parameters
           (See Eq.~\ref{nonBonding01} and ~\ref{nonBonding02} for more detail).
           O, Ti, Ba represent the atoms in the BaTiO$_3$ crystal, and
           Os and Hs denote the surface oxygen and hydrogen, respectively.
           The units for Q, eps and sigma are 
           C, kcal/mol and \AA, respectively.
     \label{TABLE03}}
     \begin{tabular}{lrrr}\hline
      &   Q     &   $\epsilon$    & $\sigma$ \\\hline 
 O    & -0.08   &  0.180   & 2.205 \\
 Ti   &  0.16   &  0.180   & 2.405 \\
 Ba   &  0.08   &  0.180   & 3.805 \\
 Os   & -0.60   &  0.482   & 1.868 \\
 Hs   &  0.60   &  0.262   & 1.300 \\\hline
     \end{tabular}
     \end{table}%

\subsection{The distribution of molecules along the z-axis near the solid interface}
\begin{figure}
          \begin{center}
           \includegraphics[clip,width=8.0cm]{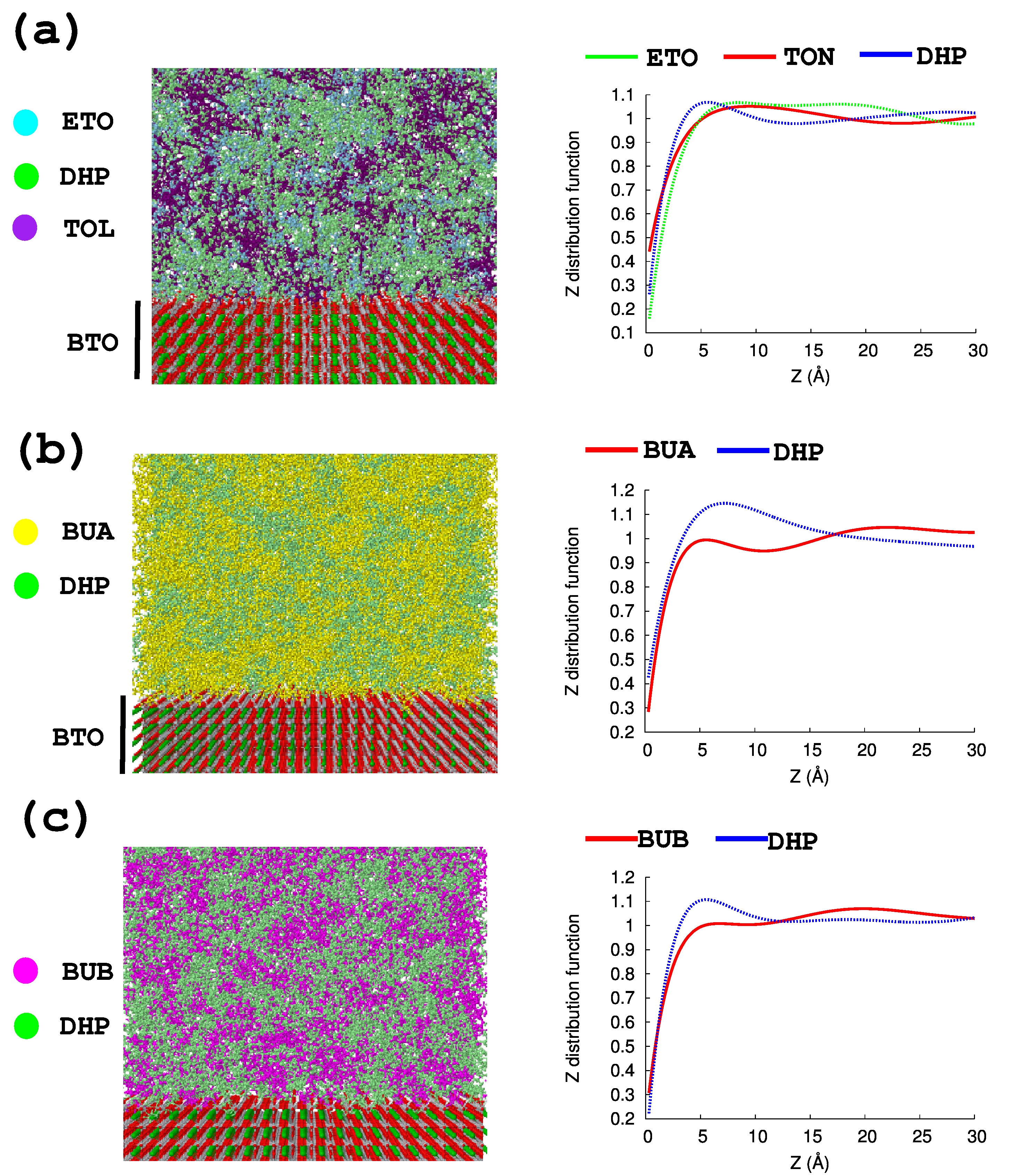}  \\
          \end{center}
	      \caption{
              (a) Distribution of ETO, TON and DHP along the z-axis,
              (b) Distribution of BUA and DHP along the z-axis,
              and (c) Distribution of BUB and DHP along the z-axis.
              \label{Figure03}
		      }
	\end{figure}

    The molecular distribution near the BaTiO$_3$ surface was investigated with the all-atom MD simulation. 
    The schematic illustration of the simulation model is shown in Figure~\ref{Figure03}-(a), 
    where the interface between the polymer molecules and the solid BaTiO$_3$ surface was generated.
    To understand the interface structure, the distribution function along the perpendicular direction was then evaluated 
    .

    The results are shown in Figure~\ref{Figure03}-(a), (b) and (c),
    where the respective figures show the distributions of (a) EOH and TON against DHP,
    (b) BUA against DHP, and (c) BUB against DHP.
    In all cases, the DHP molecules locate in the vicinity of the BaTiO$_3$ solid surface, 
    pushing other molecules to the bulk region. The most prominent distribution difference is 
    observed in the case of the mixture of BUA and DHP simulation (Figure~\ref{Figure03}-(b)), 
    because BUA contains the hydrophobic alkyl chain. 
    ETO molecules are located relatively close to the solid surface, and the 
    presence of DHP had less impact compared with other molecules, which 
    is also reasonable given the polarizability of the ETO molecule. 
     
    The all-atom MD simulation above makes clear the impact 
    of the interaction energy on the actual distribution of molecules near 
    the interface. Although the adsorption energy of DHP (35.44 kcal/mol)  
    was significantly higher than that  of the others, the probability of DHP locating 
    at a neighbor site is not high compared with that of other molecules. 
    This is because the final distributions of the molecules are determined not only 
    by the most stable interaction energy, but also by other factors such as 
    molecular size, rotation and interaction 
    between the respective molecules. 
    MD simulation is thus important for analyzing the impact of the DHP dispersant
    on the distribution of other molecules. 
        
    The simulation above is also an effective tool for screening the dispersant molecule
    for stabilizing the particles dispersion and decreasing the slurry viscosity. 
    According to Fig. 3, DHP molecules locate on the BTO surface nearer than 
    solvent (EOH and TON) and binder (BUA and BUB), indicating that DHP molecules preferentially adsorb on the BTO particles. In general, the dispersion adsorbed layer 
    provides the steric and electrostatic repulsive force, which results in
    the increase in the stability of particle dispersion. \cite{ref70,ref71}. 
    Therefore, 
    the above simulation suggests that DHP molecules are a good dispersant for BaTiO$_3$-based slurry.

\subsection{Evaluation of the rheology of BaTiO$_3$ slurry using the CG model}
    To understand the relationship between the interface geometry and the rheology of the slurry,
    a CGMD simulation was performed. The parameter for the CG model was generated using 
    the Boltzmann inversion method\cite{reith2003deriving,guerrault2004dissipative} as noted in computational detail,
    and the distribution of molecules near the interface was used to generate the pair interaction parameters
    between the solid and molecular interface.
    To validate the simulation model, we investigated the rheology for five different types of system, 
    where the composition and size of BaTiO$_3$ particles differ. The simulated composition is listed 
    in TABLE~\ref{TABLE01}, and the tested particle sizes were 2 nm (model01, model02 and model03), 
    8 nm (model04) and 15 nm (model05).

\begin{figure}
          \begin{center}
           \includegraphics[clip,width=8.0cm]{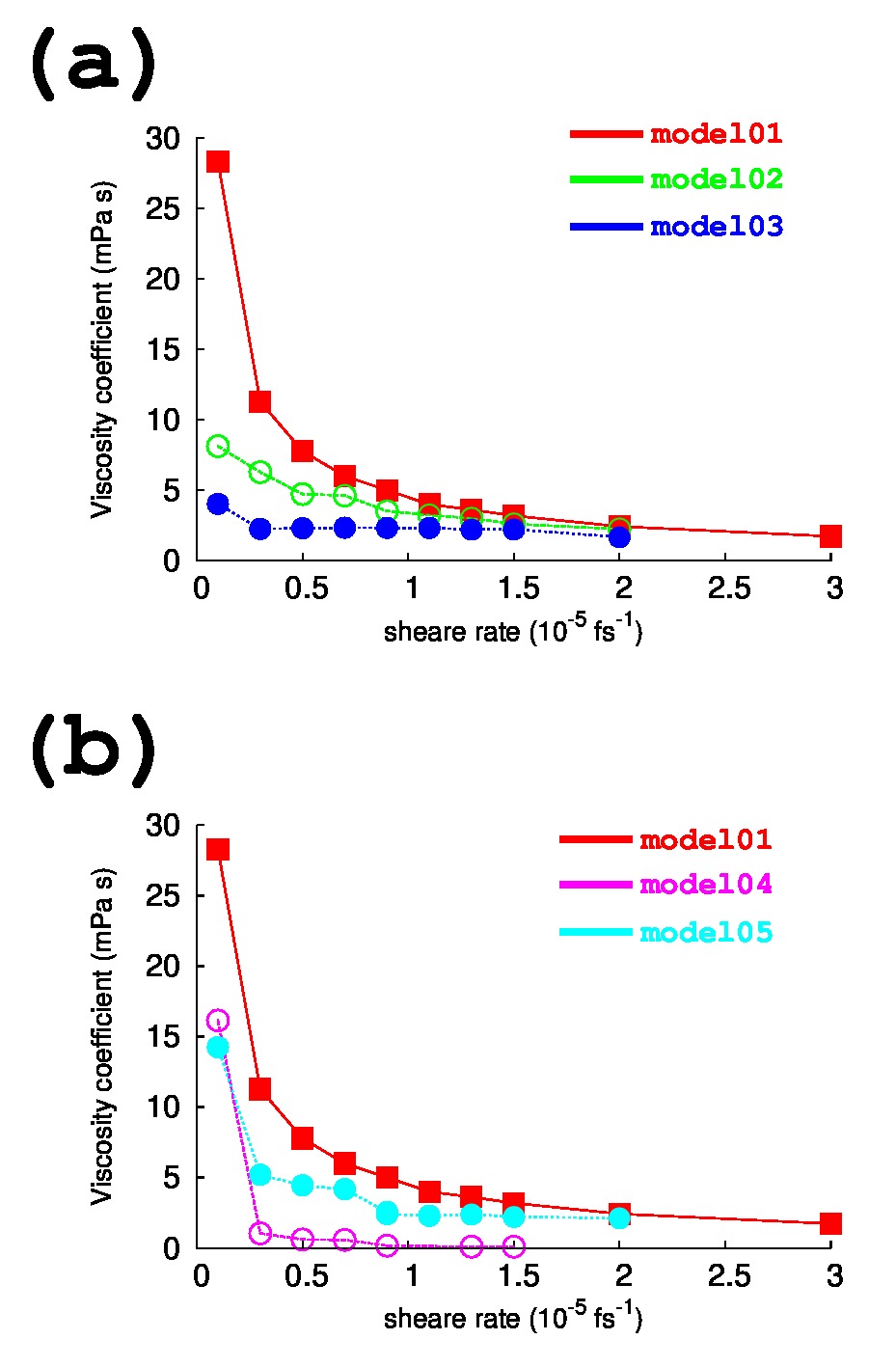} \\
          \end{center}
	      \caption{
               Viscosity analysis for BaTiO$_3$ slurry model
               (a) result of viscosity with different dispersant composition, and
               (b) result of viscosity with varying BaTiO$_3$ particle size.  
              \label{Figure04}
		      }
	\end{figure}

    The results are summarized in Figure~\ref{Figure04}. 
    The viscosity decreases as the shear rate increases in all models (Figure~\ref{Figure04}-(a)); that is, all models show shear thinning behavior. In particular, a significant decrease in viscosity with an increase in shear rate was observed in model01, resulting from aggregation of particles due to a large attractive force. In addition, the viscosity steeply decreases with the increase in DHP composition from model01 to model03.
    As predicted from the results in Figure~\ref{Figure03}, the differences in the flow behavior of the three models are clearly due to the effect of DHP as a dispersant on BaTiO$_3$.
    A decrease in viscosity with the addition of phosphate ester was also observed in 
    the experiment\cite{mikeska1988non,kim2003suspension}, and we assumed that the CG model could capture the physical picture 
    of the BaTiO$_3$ slurry. Furthermore, increasing the particle size also decreases the viscosity 
    of slurry (Figure~\ref{Figure04}-(b)), which is also a typical behavior of the BaTiO$_3$ 
    slurry\cite{shibata2012dispersion}. The simulation results indicate that the CG model used in this study can simulate the important effect of dispersant on the viscosity.

    Since the CG model could simulate the flow behavior of the BaTiO3 slurry, 
    it might be applied to know an appropriate amount of the dispersant 
    for stable dispersion of particles, which is significantly important infomation
    for manufacturing.
    In this study, three different weight fraction of DHP and BTO are investigated:
    the ratio of weight fraction is 0.1 (DHP=0.03/BTO=0.3) for moderl01, 
    0.65 (DHP=0.13/BTO=0.2) for model02, and 2.3 (DHP=0.23/BTO=0.1) for mode03. 
    In model01, the viscosity is very high at low shear rates, 
    and a large shear stress is required to change the slurry microstructure. 
    On the other hand, in the case of model02 and model03 with larger amount of dispersant
    than model01, the slurry viscosity maintains relatively low even at low shear rates. 
    Although the viscosity of slurry model3 is slightly high at a shear rate of 0.1 fs$^{-1}$,
    it behaves almost like a Newtonian fluid. 
    Therefore, we conclude that the  ratio of DHP/BTO in model03 is slightly excessive,  
    and the appropriate amount of dispersant exists between model02 and model03. 
    
    Furthermore, the optimal amount of dispersant depends on the particle size in the slurry.
    The smaller size the particles are, the larger surface area they have,
    and therefore it is necessary to add much amount of dispersant for stable dispersion of BTO particles. 
    As shown in FIG~\ref{Figure04}-(b), we have analyzed three models whose particle 
    sizes are different but same weight fraction 
    (Thus the respective models have different total surface area).
    In model04 and model05 (8nm and 15 nm particle size), 
    a sufficient amount of dispersant adsorbed on the BTO particles, 
    which results in the dispersion of BTO particles. Then the slurry 
    microstructure easily collapses due to shearing.
    Although the ratios of weight fraction is same between model01 (2nm particle size),
    model04 (8 nm particle size),
    and model05 (15 nm particle size), 
    the viscosity of model04 and model05 are significantly  lower than model01.
    Thus, the CGMD simulations suggest that the ratio of DHP/BTO=0.1 is excessive for model04 or
    model05 slurry. As mentioned above, from the results of the CGMD simulation, 
    it is possible to predict the effect of the dispersant in various slurry models,
    and thus we could predict the optimal amount of the dispersant 
    for preparation of the stable dispersion of slurry.

    To investigate the respective contribution of a particular molecule to the viscosity of the slurry, 
    we decomposed the total shear stress into the respective molecular contributions, 
    and the results are summarized in TABLE~\ref{TABLE04}.  
    The main contribution comes from the viscosity of the BaTiO$_3$ interaction site,
    Addition of the DHP dispersant drastically reduced the viscosity of the slurry model 
    from 22.44 to 2.87, suggesting that the DHP dispersant reduces 
    the stress between BaTiO$_3$ particles by locating between the surfaces. 
    A reduction in viscosity (from 5.02 to 1.20) was also observed at the solvent interaction sites,
    and the DHP dispersant also prevents adhesion of the solvent on the BaTiO$_3$ 
    surface.
     
     \begin{table}[h!]
     \caption[]{
          Viscosity for each molecular contribution.    
       \label{TABLE04}}
     \begin{tabular}{lrrrr}\hline
         & bto      &   binder   &   dispersant    & solvent    \\\hline 
 model01 &22.44     &   0.68  &  0.20        &   5.02  \\
 model02 & 5.76     &   0.21  &  0.24        &   1.16  \\
 model03 & 2.88     &   0.00  &  0.77        &   1.25  \\\hline
     \end{tabular}
     \end{table}%

\begin{figure}
          \begin{center}
           \includegraphics[clip,width=8.0cm]{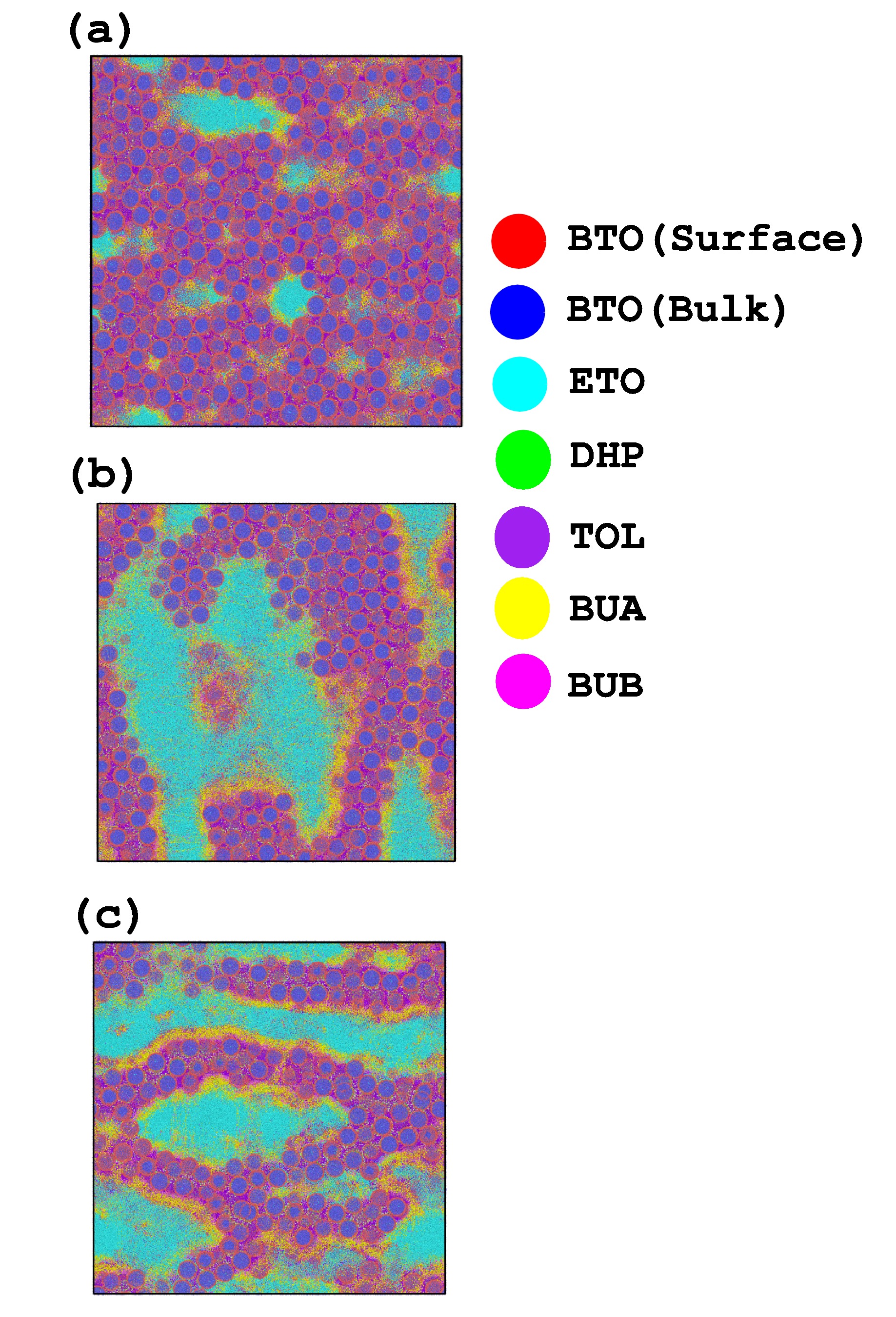} \\
          \end{center}
	      \caption{
               Final geometry of the slurry model after 5-ns MD simulation,
               (a) model01 (b) model02 and (c) model03.
              \label{Figure05}
		      }
	\end{figure}

\begin{figure}
          \begin{center}
           \includegraphics[clip,width=8.0cm]{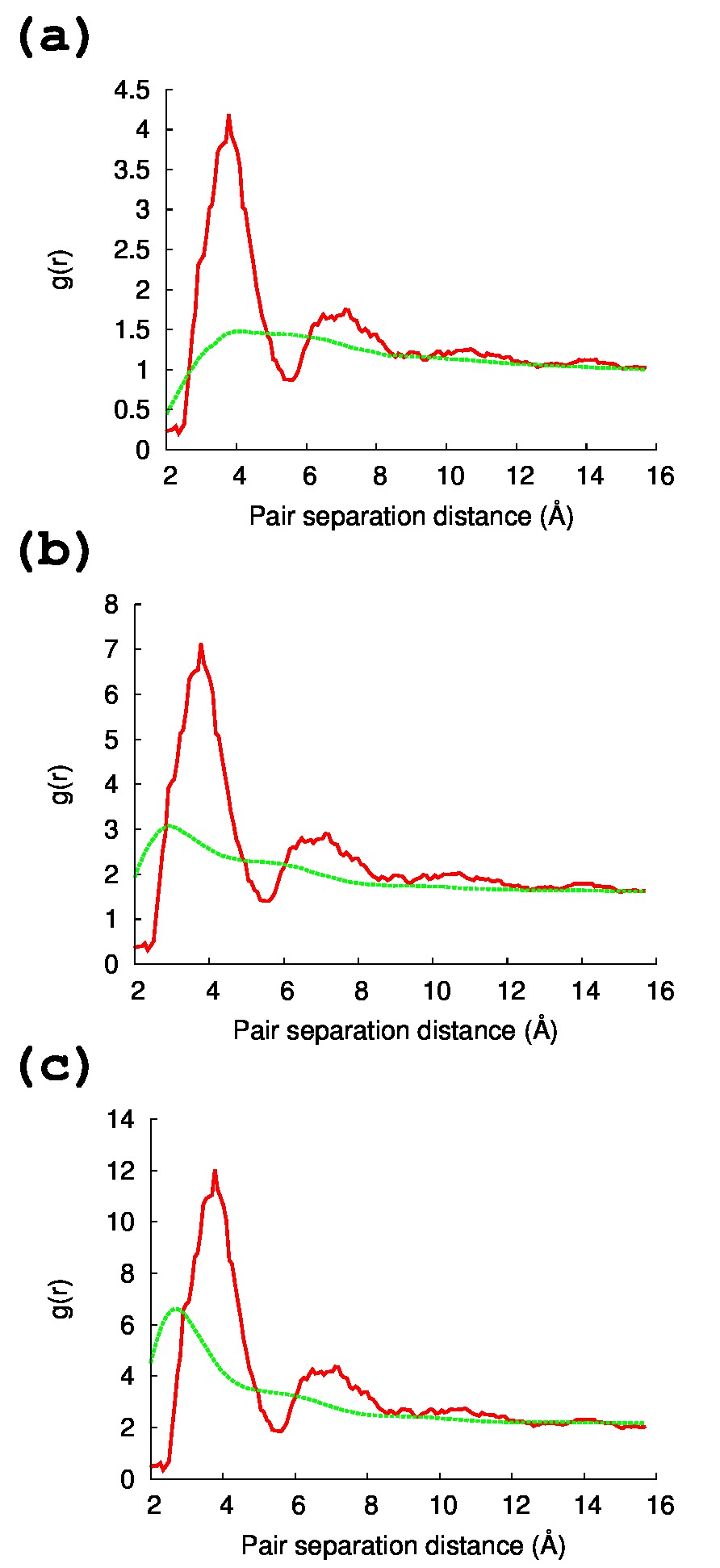} \\
          \end{center}
	      \caption{
               RDF between BaTiO$_3$ interaction sites (red solid line),
               and RDF between BaTiO$_3$ interaction sites and DHP dispersant (green dashed line) 
               (a) model01 (b) model02 and (c) model03.
              \label{Figure06}
		      }
	\end{figure}

    To understand how the dispersant affects the viscosity of the slurry,
    the structures of slurry models were investigated to determine whether BaTiO$_3$ particles 
    aggregate with each other, and where the DHP dispersant located in the slurry.
    The final structure after 5-ns simulation is shown in Figure~\ref{Figure05}.  
    In Figure~\ref{Figure05}, red (surface)  and blue (inside particle) 
    denote the BaTiO$_3$ particles, and green the DHP dispersant.
    The simulation results suggest that the BaTiO$_3$ particles aggregated in all slurry  models, 
    and no apparent difference was noted.
    To further investigate the difference between the slurry models,
    the RDF was analyzed, and the results are shown in Figure~\ref{Figure06}. 
    The red lines in Figure~\ref{Figure06} indicate the  RDF between the surface BaTiO$_3$ interaction sites,
    with no difference observed between the respective slurry models.
    By contrast, a clear difference is observed in the RDF between DHP and BaTiO$_3$ surface:
    When plenty of DHP is added (model03), the RDF shows that the DHP dispersant mainly locates
    near the BaTiO$_3$ interaction sites. The simulation results thus suggest that locating  DHP in the vicinity 
    of BaTiO$_3$ surface is a key feature in determining the rheology of BaTiO$_3$ slurry.

\begin{figure}
          \begin{center}
           \includegraphics[clip,width=8.0cm]{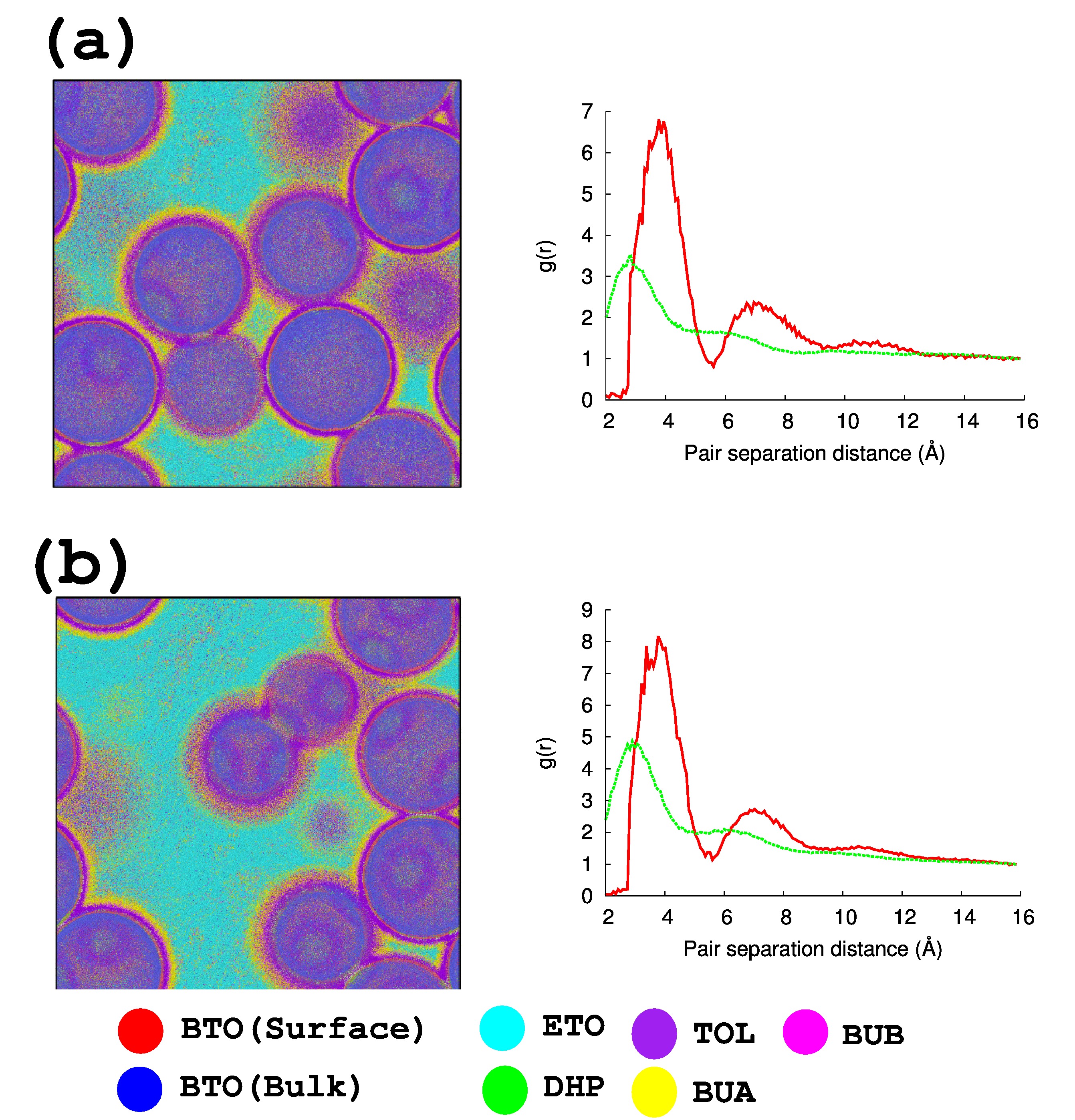} \\
          \end{center}
	      \caption{
               Final geometry of slurry model after 5-ns MD simulation (left),
               RDF between BaTiO$_3$ interaction sites (red solid line),
               and RDF between BaTiO$_3$ interaction sites and DHP dispersant (green dashed line) 
               (a) model04 and  (b) model05.
              \label{Figure07}
		      }
	\end{figure}

    To further understand the rheology of the BaTiO$_3$ slurry, 
    we also analyzed the effect of particle size (from 2 to 15 nm),
    where we also observed the reduction in viscosity with increasing particle size.
    To understand the change of viscosity, the  final structure  and 
    corresponding RDF are shown in  Figure~\ref{Figure07}. 
The RDF between BaTiO$_3$  interaction sites 
    does not change significantly (Figure~\ref{Figure07} compared with model01, but
    we observe an intense peak of the RDF between BaTiO$_3$ and DHP interaction sites, and
    the nearest-neighbor distance is less than the distance between the BaTiO$_3$ interaction sites.
    This result also suggests that the key factor for determining the viscosity of the
    BaTiO$_3$ slurry is the population of the DHP dispersant locating in the vicinity of the
    BaTiO$_3$ surface.

\section{Conclusions}
    We investigated the rheology of a BaTiO$_3$ slurry using  CGMD simulation
    to understand the role of dispersant, especially focusing on the close-to-solid surface. 
    First, the DFT and all-atom MD simulations showed that in contrast to binder or solvent molecules,
    the DHP dispersant prefers to locate near the BaTiO$_3$ surface 
    .
    By scaling up the atomistic simulation, 
    the coarse grain simulation model can then successfully reproduce the reduction in viscosity 
    with the increase of dispersant. 
    In the simulation model, the DHP dispersant preferentially located in the vicinity of the
    BaTiO$_3$ surface, playing an important role in sliding the BaTiO$_3$ particles 
    by the repulsion force, due to the hydrophobic  didodecyl group,
    suggesting that the population of dispersant near the BaTiO$_3$ surface is a key factor 
    in determining the rheology of the BaTiO$_3$ slurry.

    In this study, we proposed a means of connecting a molecular-level surface interaction 
    to a microstructure physical property such as viscosity by using quantum-level, 
    all-atom MD and CGMD simulations.
    By scaling up from the atomistic-level simulation to CG simulation, we were able to grasp in a rigorous manner how the adsorption energy difference between respective molecules
    results in a final microstructure property such as rheology.
    Such a molecular-level understanding of the BaTiO$_3$ slurry enables the prediction of slurry macro behavior 
    and the estimation of optimal additive amount of dispersant 
    without experiments, and thus could be of great help in the efficient optimization of slurry preparation.

\section*{ACKNOWLEDGMENTS}
We thank 
the Research Institute for Information Technology
 at Kyushu University for providing computational resources. 
This research also used the computational resources of 
the Fujitsu PRIMERGY CX400M1/CX2550M5(Oakbridge-CX) at the
Information Technology Center of The University of Tokyo
through the HPCI System Research project (Project ID:hp200015).


\newpage

\end{document}